# Distributed Transformer Monitoring System Based On Zigbee Technology

Rakesh Kumar Pandey [#1] Dilip Kumar [*2]

[1-2]Academic and Consultancy Services Division,

Centre for Development of Advanced Computing (CDAC), Mohali, India

*Abstract*— **A distributed transformer networks remote monitoring system(DTRMS) is developed and constructed, for monitor and record the parameters like temperature, oil level status, of a distribution transformer. The system consists of a microcontroller based circuit, with solid-state components for handling sensors, power back-up, real time clock and data communication module which based on ZigBee protocol. Sensors, including a Resistance Temperature Detector (RTD) and a Liquid level sensor performs according to manufacturers' specifications are calibrated and tested using LabVIEW software. The system is installed at the distribution transformer site and by measuring above parameters it will help the utilities to optimally utilize transformers and identify problems before any catastrophic failure.**

*Keywords*— **DTRMS; Distribution Transformers, RTDs, Level Sensor, ZigBee**

## I. INTRODUCTION

Transformer is the key equipment in power system, to ensure its safe and stable operation is important. Transformers either raise a voltage to decrease losses, or decreases voltage to a safe level. "Monitoring" is here defined as on-line collection of data and includes sensor development, measurement techniques for on-line applications. It is very difficult and expensive to construct the communication wires to monitor and control each distribution transformer station. Here ZigBee is used for communicating the monitored parameters.

The failures of transformers in service are broadly due to: temperature rise, low oil levels, over load, poor quality of LT cables, and improper installation and maintenance. Out of these factors temperature rise, low oil levels and over load, need continuous monitoring to save transformer life.

A DTRMS increases the reliability of distribution network, by monitoring critical information such as oil temperature, and oil level of transformer. Data are collected continuously. Monitoring the transformers for problems before they occur can prevent faults that are costly to fix and result in a loss of service life.

Although there are other monitoring systems available, including Dissolved Gas Analysis (DGA) and Particle Discharge (PD) [1]. C. Bengtsson has discussed Status and Trends in Transformer Monitoring [2]. A survey of the most important methods for on-line monitoring and off-line diagnostics is given in this paper. Abdul-Rahman AI-Ali et al., 2004 has used GSM to monitor distributed transformer [3].

Monitoring system based on ZigBee technology that has the potential to be a more accurate and cheaper technique for health assessment of transformers has been developed and presented in this paper. ZigBee is a specification for a suite of high level communication protocols using small, low-power digital radios based on the IEEE 802.15.4-2003 standard for wireless personal area networks [4]-[5].

## II. General description of the DTRMS system

The DTRMS is divided into two sections:
1. Coordinator
2. End device

The end device contains different types of sensors. The sensors are used to sense the different parameters and send it to microcontroller. Microcontroller transmits through ZigBee transmitter.

The Coordinator receives this data and displays it on the LCD. This whole process is real time monitoring parameters of the transformer.

It is installed at the distribution transformer site and the parameters are recorded using the built-in 8-channel analog to digital converter (ADC) of the microcontroller. Sensors are used to sense the different parameters and send it to microcontroller. The acquired parameters are processed and recorded in the system memory and transmitted to coordinator unit. On the other hand, the coordinator receives this data and displays it on the LCD.

DTRMS system has outstanding three advantages: (1) it can allow for a change from periodic-to condition based maintenance. (2) ad-hoc communication network (3) by monitoring the important functions of the transformer, developing faults can be detected before they lead to a catastrophic failure. The main benefits for WDTMS technology is low installed cost, less time for installation, safe operation and more reliable service.

*A. Circuit design*

AVR Microcontroller, Temperature sensor (RTDs), Liquid level sensor, Voltage regulators, Operational amplifier, LCD Display and ZigBee are the main component of DTRMS.

The block diagram for ZigBee sensor network is shown in figure1.

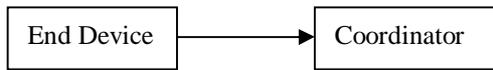

**Fig.1** Block Diagram of Zigbee sensor network

**(a) End Device**:

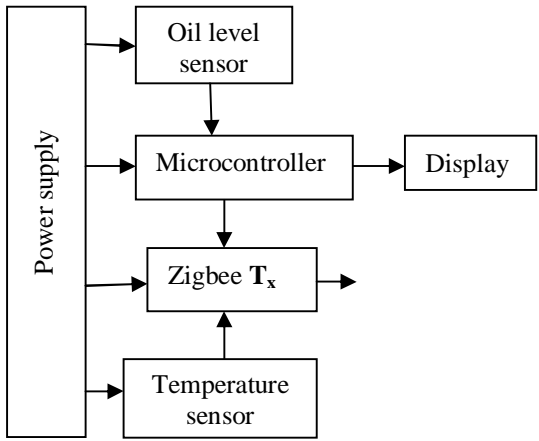

**Fig.2.** Block diagram of the end device

End device is considered to be a transmitter. The main components of End device are a programmable microcontroller, non-volatile memory (RAM), voltage regulators, Op-Amp, Temperature sensor (RTDs), Liquid level sensor, and LCD. End device consists of power supply which provides 5V DC to sensors and MCU. The power supply also produces 3.3V DC for ZigBee module & +15V,-15V DC for Operational amplifier circuit. The condition of the oil level recorded through sensor is send to ZigBee module for transmission using ZigBee protocol at 2.4 GHz. Temperature sensor attached to the In-Built ADC of the ZigBee gives the data in digital frame for the direct transmission. The microcontroller decides the condition of transformer according to the predefined threshold value. The Basic block diagram of the end device is shown in figure 2.

**(b) Coordinator:**

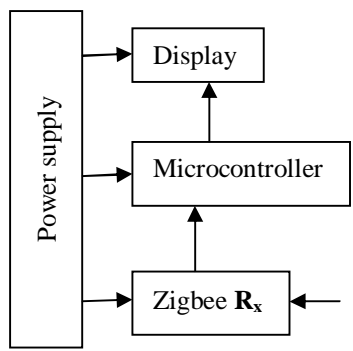

**Fig.3.** Block diagram of the coordinator unit

The coordinator also consists of power supply which produces 5V for MCU and LCD and 3.3V for ZigBee module. The ZigBee module receives digital data from end device and transfers it to microcontroller serially. Then microcontroller displays the condition of oil and temperature of transformer. Figure 3 shows the basic block diagram of coordinator unit.

III. CIRCUIT OPERATION

Temperature sensor (RTD) is used for top oil temperature measurement and oil level sensor is used for measuring the conservator oil level of the transformer. Full-bridge circuit is used to convert temperature sensor reading to a compatible signals that can be read by the microcontroller built-in ADCs (0-5 volts DC). A set of resistors are used to adjust the gain and the offset of Op-Amp. A set of rectifier circuits and center taped transformer is used to convert and scale the current and voltage values to compatible levels with the Op-Amps circuits.

The microcontroller has 8 channels, 10-bit analog-to-digital converter. ADC is used to read the parameters of sensors. A display unit, which may be an LCD display that receives display signals from the microcontroller and displays the parameters of transformer. Parameters are also transmitted to coordinator unit through Zigbee module. On the other hand coordinator unit has also a Zigbee module for receiving data from transmitted unit. After receiving transmitted data by end device, coordinator unit displays it through microcontroller. The Schematic of hardware is shown in figure 4.

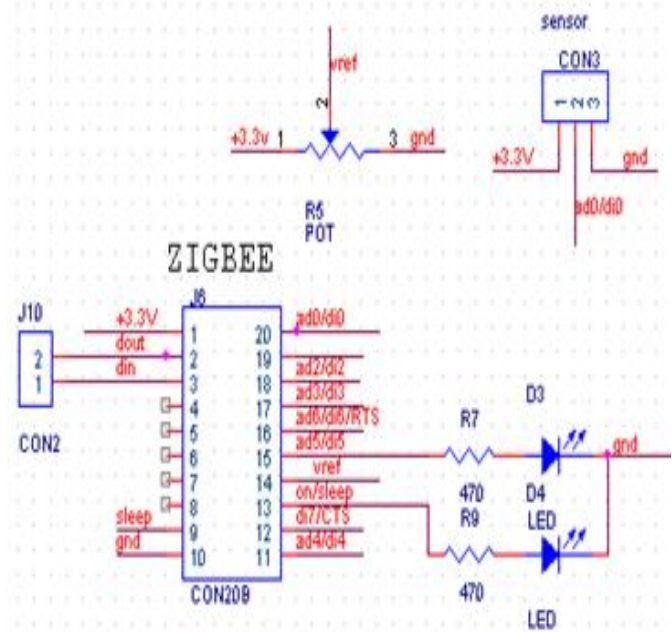

**Fig.4.** Schematic of hardware

## IV. SOFTWARE DESIGN

The software of the DTRMS is composed of the initialization and data communication. Initialization includes initialization of the input/output ports, data direction flow, set the ADC channels and reset all related memory locations that are going to be used in the operation. Data communication means transfer of measured data from end device to coordinator unit. X-CTU is a Windows-based application provided by Digi used for programming the Zigbee module as shown in figure 5.

The voltage at the transmitter end is compared with the digital data getting at the receiver end. This mean received data & voltage provided by the sensor for different emperature are always in a same ratio.

Once the ratio between the sensor output voltage and the receive data are same, received data is displayed on the LCD screen with the help of the microcontroller.

Now change the temperature at the transmitter end by heating up the sensor, check the change in the temperature at the receiver end.

Now change the temperature of other xbee sensor & get the reading on receiver end.

## VI. CONCLUSION

All the objectives outlined in this paper are achieved. The study of ZigBee modules available in market was done and the best ZigBee module was chosen. The chosen modules were studied and were implemented as end device and coordinator. Successful communication was setup between coordinator and end device. There are two most important feature of this product. First one is the use of ZigBee technique to transfer data from one point to other; this method increases the life of battery and the product. Using this technology it's possible to cover large fields of about 1 km square area.

With modern technology it is possible to monitor a large number of parameters of distributed transformer at a relatively high cost. The challenge is to balance the functions of the monitoring system and its cost and reliability. In order to get effective transformer monitoring system to a moderate cost, it is necessary to focus on a few key parameters. WDTMS is able to record and send abnormal parameters of a transformer to concerned office. It works on Zigbee technology that supports multiple network topologies such as point-to-point, point-to-multipoint and mesh networks. It has low duty cycle – provides long battery life.

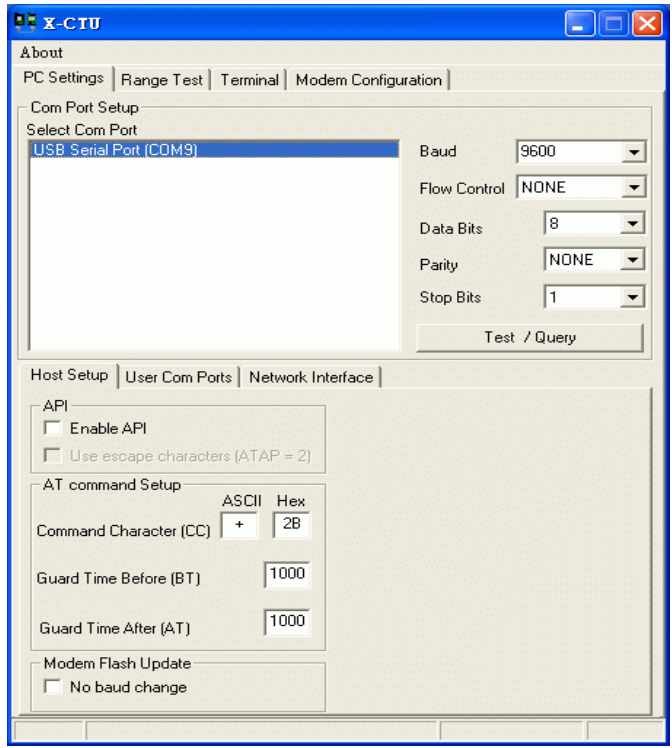

**Fig.5.** X-CTU Software

### V. TESTING RESULTS

Prior to deployment on actual condition, it is necessary to test the circuit module by module, to make sure that the complete circuit is working properly. So first the ZigBee module, then the microcontroller circuit and then Sensors were tested and finally the compete circuit were tested.

The complete project is tested in following steps:
First the temperature sensor is calibrated and tested by using LabVIEW software to examine the accuracy and variability of the sensor measurements. The temperature sensor RTD is placed in a controlled-temperature environment, and calibrated for the different temperature by taking the voltage reading of the sensor at the transmitter end.